\documentstyle[preprint,aps]{revtex}

\begin{document}
\title{Sources of CMB Spots in Closed Hyperbolic Universes}
\author{Helio V. Fagundes}
\address{Instituto de F\'{i}sica Te\'{o}rica, Universidade Estadual Paulista,\\
Rua Pamplona, 145, 01405-900 S\~{a}o Paulo, SP, Brazil}
\date{\today}
\maketitle

\begin{abstract}
The results of a previous paper by the author [Astrophys. J. 470, 43 (1996)]
are improved, through a more reliable technique for finding the location of
sources whose images are on the surface of last scattering.
\end{abstract}

\pacs{98....}

\section{\protect\bigskip INTRODUCTION \ }

Some years ago Cay\'{o}n and Smoot \cite{CS}; hereafter CS, identified a
number of \ `cold' and `hot' spots in the COBE maps of the cosmic microwave
background (CMB), as patches of physical density fluctuations (rather than
noise) on the surface of last scattering (SLS). A cold (hot) spot,
interpreted as gravitational Sachs-Wolfe effect plus fluctuation of
radiation temperature on the SLS, corresponds to an increase (decrease) of
matter density \cite{White}.\ These results were used by the author \cite
{ApJ470(96)43}; henceforth Paper I, in connection with the possibility of
the universe's spatial section being a closed hyperbolic 3-manifold (CHM).

Tables 1a-b list the galactic coordinates in degrees $(-180<l\leq 180,\
-90\leq b\leq 90)$ for the positions of the six overerdense $(C1-C6)$ and
the eight underdense $(H1-H8)$ CS spots. Given the comoving nature of cosmic
geometry, it was argued that those spots, for example the overdense ones,
might have evolved into galaxy superclusters which are - or may become -
observable in our epoch. The underdense spots would have evolved into the
relative voids in the observed structure of the large-scale matter
distribution. \ 

\qquad \qquad \qquad {\bf \qquad\ \ \ \ \ \ \ \ \ \ \ \ \ \ \ \ \ \ \ \ \ \ }%
\qquad

\qquad 
\begin{tabular}[t]{cccr}
\multicolumn{4}{c}{\bf Table 1a. Cold spots} \\ \hline\hline
Spot &  & $l$ & $b$ \\ \hline
$C1$ &  & \multicolumn{1}{r}{$-99.0$} & $57.0$ \\ 
$C2$ &  & \multicolumn{1}{r}{$85.0$} & $40.0$ \\ 
$C3$ &  & \multicolumn{1}{r}{$-21.0$} & $-45.0$ \\ 
$C4$ &  & \multicolumn{1}{r}{$73.0$} & $-29.0$ \\ 
$C5$ &  & \multicolumn{1}{r}{$81.5$} & $-59.0$ \\ 
$C6$ &  & \multicolumn{1}{r}{$-86.0$} & $33.0$ \\ \hline
\end{tabular}
\qquad \qquad \qquad \qquad 
\begin{tabular}[t]{cccr}
\multicolumn{4}{c}{\bf Table 1b. Hot spots} \\ \hline\hline
Spot &  & $l$ & $b$ \\ \hline
$H1$ &  & \multicolumn{1}{r}{$-24.0$} & $51.0$ \\ 
$H2$ &  & \multicolumn{1}{r}{$46.0$} & $-32.0$ \\ 
$H3$ &  & \multicolumn{1}{r}{$96.0$} & $-28.0$ \\ 
$H4$ &  & \multicolumn{1}{r}{$120.0$} & $-36.6$ \\ 
$H5$ &  & \multicolumn{1}{r}{$172.0$} & $29.0$ \\ 
$H6$ &  & \multicolumn{1}{r}{$-81.0$} & $-33.0$ \\ 
$H7$ &  & \multicolumn{1}{r}{$141.0$} & $-73.5$ \\ 
$H8$ &  & \multicolumn{1}{l}{$-82.0$} & $-58.0$ \\ \hline
\end{tabular}

\ 

Since the CMB spots are the result of using the COBE satellite's wide-angle
detectors, they represent a smearing of density inhomogeneities on the scale
of about ten arc degrees. But we may expect that the small angle detectors
in the planned projects MAP and Planck will produce spots that cluster in a
way consistent with COBE's maps. In the bottom-up model of structure
formation \cite{Peebles} these fine-grained overdense spots would evolve
into galaxies, which would later produce clusters and superclusters.
Therefore the original idea that COBE spots become superclusters (or
large-scale voids) seems basically sound.

The purpose of Paper I was to fit a number of CHM's to the CS spots. Since
these are interpreted as density inhomogeneities in the fundamental
polyhedron (FP) for the manifold, we hoped the positions of the latter, when
compared with those of observed structures and voids, might favor some of
those CHM's as the real cosmic space. This task is of course far from
simple: there are infinitely many CHM's, Earth might be located anywhere
inside them, with any orientation with respect to the real sky, not to
mention the uncertainties about $\Omega _{0}$, $H_{0}$, and now also $%
\Lambda $! But we may expect this situation to improve: the parameters'
ranges will be narrowed, and the several proposals for topology
determination (cf. Weeks's \cite{JW} and Roukema's \cite{Roukema} talks in
this Session) will eventually indicate what manifolds are the most likely
candidates for representing the cosmos.

\bigskip

\section{THE SEARCH FOR SEEDS IN THE FUNDAMENTAL POLYHEDRON}

In Paper I a lopsided method was adopted to choose a possible source for a
given CS spot. The sources shown in Fig. 3 and Tables 2 and 3 of Paper I are
usually concentrated in narrow bands of galactic latitude $b$. Another
problem with Paper I was that we looked for sources inside the maximum
injectivity FP, with basepoint at the center (which is the standard in
SnapPea \cite{SnapPea}) and the observer displaced from the center; this
produces an asymmetry in the distances of the found candidate sources.\qquad

Here I make two improvements on this research, both with the help of our
geometer guru, Jeff Weeks. One of them is that now $FP$ is chosen with
basepoint on the observer's position, which makes the source distribution
centered on the the observer. The other one is in the search procedure,
which will now be discussed.

Fig. 1 is a crude sketch of $FP$ for a CHM isometric to a quotient space $%
H^{3}/\Gamma $ as usual, with the basepoint and observer at $FP$'s center $O$%
, and the spherical $SLS$.

With a point $P^{\prime }$ (the center of a spot) $\in $ $SLS$ we want to
find the pre-image $P$ of $P^{\prime }$ in $FP$, that is, $P^{\prime
}=\gamma P$ for some $\gamma \in \Gamma $. Let $\{g_{k},\ k=1-n\}$ be the
set of face-pairing generators corresponding to the $n$-face $FP.$ Points $P$
and $P^{\prime }$, and cell $FP^{\prime }=\gamma (FP)$ are also shown in
Fig. 1.

The computer search for $P$ and $\gamma $ is shown in Fig. 2 as a
flow chart, where $d(A,B)$ means comoving distance between
points $A$ and $B$; $g$ is represented by G, $g_{k}$ by Gk, and inv(G) is 
the inverse of $g$. Referring to this figure, note that (i) if the result is 
$\gamma =g=1$, then $P=P^{\prime }\in FP$, which means that $FP$ is large
enough to intersect $SLS$; and (ii) the calculation of $\gamma $ is not
necessary, but is useful for checking: it must be $P^{\prime }=\gamma P$.

The search for sources was done in Paper I for the ten smallest known CHM's,
numbered HW1-10; they are described in \cite{HW}. For each of these the
observer's position and space orientation were chosen randomly, but, as said
above, not so randomly distributed sources were found. Here we get better
results, as shown for manifold $m007(+3,1)$ or HW3 in Table 2. The observer
is at position $(0.3,\ 0.0,\ 0.0)$ in Klein coordinates relative to the
standard $FP$ in SnapPea, and axes rotated by $(4.5669,0.1078,5.3451)$ in
Euler angles in radians, with respect to the axes obtained for the displaced
basepoint. (It is interesting to observe that the standard $FP$ has 20
faces, against 24 faces in our example.) Table 2 shows $(Z,l,b)$ and the
path $g=\gamma ^{-1}$ from the spot to its source, as a word in the
generators ($g_{1}-g_{24}$ are here labeled $G00-G23$, as in the computer
printout). We adopt the values $\Omega _{0}=0.3$ and $Z_{SLS}=1300.${\bf \ }

The idea to pursue, which is illustrated in Paper I, is to vary these
positions and orientations randomly, until we get sources that match data in
catalogs of galaxy superclusters and voids - which, for this to become
possible, should reach much deeper space than the present limit of about $%
Z=0.12$ in Einasto et al.'s \cite{Einasto} list of superclusters.\qquad
\qquad \qquad

\qquad\ \qquad\ \qquad\ \ \qquad \qquad\ \ \ \ \ \ \ \ \ \ \ 
\begin{tabular}{crrrrc}
\multicolumn{6}{c}{\bf Table 2. Sources for CS spots in m007(+3,1) universe.}
\\ \hline\hline
Spot &  & \multicolumn{3}{c}{
\begin{tabular}{lll}
& Source &  \\ \hline
$Z$ & \multicolumn{1}{c}{$l\ \ $} & \multicolumn{1}{r}{$b$}
\end{tabular}
} & Motion from spot to source \\ \hline\hline
$C1$ &  & $1.133$ & $-51.3$ & $12.0$ & G01G12G08G06 \\ 
$C2$ &  & $0.638$ & $-94.4$ & $79.4$ & G07G14G06G04G01 \\ 
$C2$ &  & $0.490$ & $63.4$ & $-29.9$ & G00G20G13G06G03 \\ 
$C4$ &  & $1.370$ & $145.0$ & $-22.4$ & G00G04G02G01 \\ 
$C5$ &  & $1.294$ & $177.7$ & $4.8$ & G06G23G04G02G21G07G05G01 \\ 
$C6$ &  & $0.477$ & $-142.1$ & $-50.0$ & G05G18G06G03G00 \\ 
&  &  &  &  &  \\ 
$H1$ &  & $1.026$ & $42.0$ & $-36.3$ & G02G21G13G05 \\ 
$H2$ &  & $0.925$ & $-91.3$ & $52.6$ & G10G07G04G02G01 \\ 
$H3$ &  & $0.564$ & $-62.8$ & $35.0$ & G01G15G10G02G01 \\ 
$H4$ &  & $1.158$ & $31.5$ & $5.3$ & G04G01G21G14G03G01 \\ 
$H5$ &  & $0.931$ & $28.3$ & $-0.1$ & G02G01G13G02 \\ 
$H6$ &  & $1.027$ & $-158.8$ & $21.4$ & G10G02G00G17G06G00 \\ 
$H7$ &  & $0.917$ & $90.8$ & $-54.4$ & G03G18G06G04 \\ 
$H8$ &  & $0.361$ & $128.1$ & $-37.6$ & G06G04G02G00 \\ \hline\hline
\end{tabular}
\ \ \ \ 

\bigskip

\ \ \ \ \ \ \ \ \ \ \ \ 

ACKNOWLEDGEMENTS

I am grateful to Funda\c{c}\~{a}o de Amparo \`{a} Pesquisa do Estado de
S\~{a}o Paulo (FAPESP-Brazil)\ for a grant to allow my participation in this
Meeting. Also to Jeff Weeks for his continued assistance on geometry
matters, and to Laerte Sodr\'{e} Jr. for a conversation on superclusters. \
\

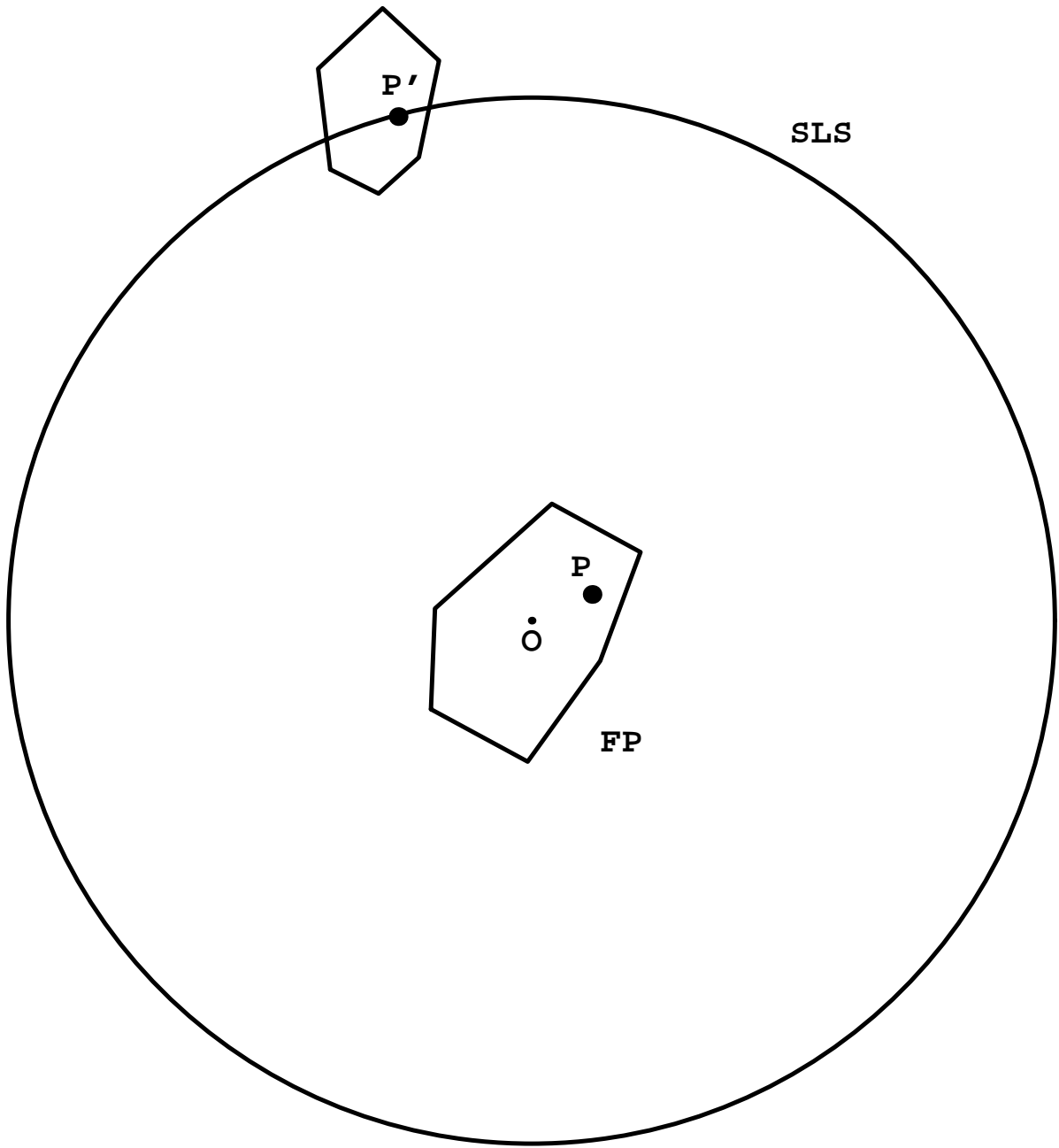
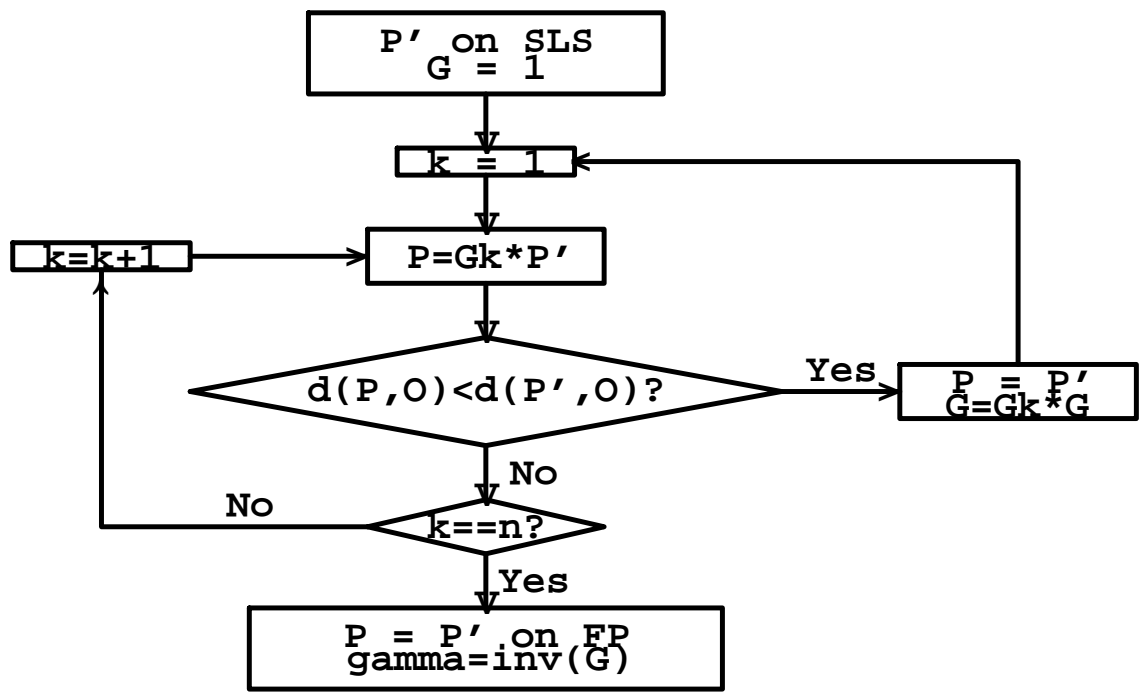
\end{document}